\documentclass[MSNbibl,nameyear,dvips]{arxstspdf}
\usepackage{flushend}
\usepackage{stfloats}

\volume{29}
\issue{1}
\pubyear{2014}
\firstpage{101}
\lastpage{102}
\doi{10.1214/13-STS447} 
\referstodoi{10.1214/13-STS420}

\begin{document}
\begin{frontmatter}
\vspace*{6pt}
\title{Discussion}
\runtitle{Discussion}

\begin{aug}
\author[a]{\fnms{A. H.} \snm{Welsh}\corref{}\ead[label=e1]{Alan.Welsh@anu.edu.au}}
\runauthor{A. H. Welsh}

\affiliation{The Australian National University}

\address[a]{A. H. Welsh is E. J. Hannan Professor of Statistics,
The Australian National University,
Canberra ACT 0200, Australia \printead{e1}.}

\end{aug}



\end{frontmatter}

This collection of papers gathering and promoting highly successful
applications of Statistics is a good antidote for anyone feeling
somewhat defensive about Statistics. The focus on the successful use of
Bayesian methods has produced a powerful and stimulating set of
stories; the Editors and Authors are all to be congratulated on their
successful efforts to bring out the stories behind these analyses. The
papers are relatively short (as was required by the Editors) and a good
measure of their success is that they both stand alone and motivate the
reader to follow up and read the original papers.

The article on the search for the wreckage of flight AF 447 (Stone et
al.) is fascinating. The description of the careful and detailed
thinking about what might have happened, the evaluation and inclusion
of relevant empirical evidence to quantify the possible scenarios and
the final success of the analysis in assigning substantial posterior
probability to where the wreckage was ultimately found are all
inspiring. Like many inspiring articles, it challenges us to think
about both the difficult issues of the particular problem considered
and general issues about the overall approach. I think a Bayesian
analysis is highly appropriate for this problem, but it is not so easy
to explain why and it is clear that, as always, the analysis itself has
to be done extremely well.

One motivation for doing a Bayesian analysis for this problem (and one
that is commonly articulated) is that the event in question is unique
so it is not meaningful to think about replications. This is not really
convincing because hypothetical replications are hypothetical whether
they are conceived of for an event that is extremely rare (and in the
extreme happens once) or for events that occur frequently. Moreover, it
turns out later that nine past crashes were deemed similar enough to be
used to provide information for constructing the prior, making it
difficult to argue that the event really is unique.

Another widely used motivation for Bayesian analysis is that it
propagates the uncertainty correctly. This is true and important, but
it is also true that it propagates only the uncertainties that we
decide to include in the model. We make choices over what uncertainties
to include and we also make relatively arbitrary choices which we
subsequently treat as fixed. For example, were the uncertainties in the
weights for the different scenarios or the chosen $\alpha$ propagated
through to the conclusion? As a practical matter, I do not believe we
can or should try to propagate all uncertainty, simply that we should
not get too carried away and forget about aspects we have treated as
certain. This highlights the fact that the Bayesian approach is a tool
that is extremely useful for combining the quantitative information we
choose to use and are able to express in terms of distributions but
which, like any tool, needs to be used well to be effective; the tool
on its own does not solve the problem but needs to be applied by highly
skilled people.

The four unsuccessful searches that preceded the final, successful
search highlight some of the issues. They too used assumptions and
information to select the search location. Presumably they did not use
a Bayesian analysis? If they did not (and it is not really possible
with the benefit of hindsight to go back and redo this fairly),
differences between the particular techniques used may be outweighed by
differences in the information and beliefs that fed into the analysis.
For example, the fourth search based on possible drift concentrated in
a small rectangle relatively far from the actual crash site. Would a
Bayesian analysis based on the information used to come up with that
search rectangle have produced different results? It is difficult to be
sure from the maps but it looks like a passive acoustic search actually
covered the crash site but that the wreckage was not discovered. We can
interpret this as measurement error or as using an incorrect prior. The
searchers tried to find the sonar beacons, not realizing that these had
failed and were not operating. The successful search both allowed for
this possibility (at least by not ruling out that area as having been
previously searched) and, because so much time had elapsed that the
beacons could not have been expected to still work, adopted different
technology in the search. Had they adopted the belief that the area had
been searched so the wreckage could not be there and built this into
the prior, it would not have been found. Thus, it was crucial to adopt
the correct beliefs to end up with the right result. The point is that
the tool had to be used well and the credit is due to the users rather
than simply the tool.

In their very interesting paper on managing Baltic salmon, Kuikka et
al. make the point that Bayesian methods make it possible to combine
``relevant data from many sources.'' The paper explicitly acknowledges
the role of politics in salmon management and the need to combine
empirical data with ``data'' that is too difficult or expensive to ever
be collected. The word relevant is critical here since irrelevant data
may at best just increase uncertainty and at worst lead to seriously
wrong answers. The choice of what is relevant or not depends ultimately
on the user and is not an automatic property of the approach. Kuikka et
al. also make the point that biologically realistic models for salmon
involve too many parameters to fit without using informative priors.
This is mentioned again in Carroll's \mbox{intriguing} paper on dietary
consumption; Bayesian computations can be used to fit models that
frequentist methods cannot fit. Running a Bayesian computation will
produce numbers but, as in any computation, we need to convince
ourselves that the numbers are meaningful before we use and interpret
them. In particular, it is important to understand clearly whether the
model is identifiable or not, whether the model is incorrect in some
important way (so the computational issues reflect lack of fit) and the
extent to which the prior is driving the analysis. The fact that these
questions are not easy to answer with complicated models and
high-dimensional parameter spaces does not lessen the importance of
trying. Identifiability is important because it is resolved by using
informative priors which regularize the likelihood and enable the model
to be fitted; even vague priors can be informative in this context.
There is no problem with using informative priors but we need to know
when the priors are informative, particularly when they are so
informative that the posterior is essentially the prior. Conceptually,
this may not be so different from the frequentist approach of imposing
nonestimable constraints on the parameters. A different kind of
identifiability issue arises in Bayesian history matching (Vernon et
al.) because it is possible that different scenarios or models can lead
to the same observable data, particularly when this is a single slice
in time. Here, finding matching simulations seems only part of the
really difficult scientific problem being considered.

Another reason a model may be difficult to fit is that it does not
describe the data. Forcing it to ``fit,'' for example,
by switching to a Bayesian analysis, may not be the best response. It
is difficult to check complicated models, particularly hierarchical
models with latent variables, measurement error, missing data, etc.,
but using an incorrect model may be a concern when the model proves
difficult to fit.

A challenging issue acknowledged in Carroll is the issue of using
survey weights in a Bayesian analysis. We can think about this as a way
of estimating the likelihood by the pseudo-likelihood and then using
this estimated likelihood in a regular Bayesian analysis. This does
involve a combination of design-based and model-based approaches which
require different conditioning but, somewhat like approximate Bayesian
computation (ABC), it might be viewed as a pragmatic approach to
solving difficult problems. It is not clear what the Bayesian costs and
benefits are; in frequentist analysis, Chambers et al. (\citeyear{Chaetal12}) show that
pseudo-likelihood estimation is less efficient than maximum likelihood
estimation so there is some loss of information. Constructing the
likelihood requires including all the design variables in the model.
Aside from the fact that, in contrast to the survey weights, the design
variables are not usually available to secondary analysts, the study
from which the data are taken (NHANES) uses a complicated design (with
several nested levels of cluster sampling) which it would not be
straightforward to incorporate into the model. Moreover, making the
model more complicated may increase the computational difficulties of
fitting the model. The use of pseudo-likelihood in Bayesian analysis
definitely needs research into its meaning and consequences before we
can consider it with equanimity.\looseness=1



\renewcommand\bibname{Reference}
%

\end{document}